\documentclass[preprint]{aastex701}

\usepackage{amsmath}
\usepackage{bm}
\usepackage{verbatim}
\usepackage[normalem]{ulem}

\shorttitle{accretion time}
\shortauthors{Li et al.}

\begin{document}

\title{Estimating accretion times of halo substructures in the Milky Way}

\author[0000-0002-9497-8127]{Hefan Li}
\affiliation{National Astronomical Observatories, Chinese Academy of Sciences, Beijing 100101, People's Republic of China}
\affiliation{Astronomical Institute, Tohoku University, 6-3 Aoba, Aramaki, Aoba-ku, Sendai, Miyagi 980-8578, Japan}
\email[show]{lihf@bao.ac.cn}

\author[0000-0002-9053-860X]{Masashi Chiba}
\affiliation{Astronomical Institute, Tohoku University, 6-3 Aoba, Aramaki, Aoba-ku, Sendai, Miyagi 980-8578, Japan}
\email[show]{chiba@astr.tohoku.ac.jp}

\author[0000-0002-0642-5689]{Xiang-Xiang Xue}
\affiliation{National Astronomical Observatories, Chinese Academy of Sciences, Beijing 100101, People's Republic of China}
\affiliation{Institute for Frontiers in Astronomy and Astrophysics, Beijing Normal University, Beijing 102206, People's Republic of China}
\email{xuexx@nao.cas.cn}

\author[0000-0002-8980-945X]{Gang Zhao}
\affiliation{National Astronomical Observatories, Chinese Academy of Sciences, Beijing 100101, People's Republic of China}
\affiliation{School of Astronomy and Space Science, University of Chinese Academy of Sciences, Beijing 100049, People's Republic of China}
\email{gzhao@nao.cas.cn}

\begin{abstract}
\par To unravel the formation history of the Milky Way, we estimate the accretion times of six phase-space substructures in the stellar halo, using the orbital frequencies toward two spatial directions ($r, \phi$) in spherical coordinates. These substructures, identified in our previous studies, are located in the solar neighbourhood and therefore have high-precision astrometry from Gaia. The uncertainties of the results are determined using the Monte Carlo method, and the significance is established through comparison with random halo samples. The results for the substructure GL-1 in both directions show good consistency and high significance ($4.3\sigma$ and $3.9\sigma$), yielding a combined accretion time of $5.6 \pm 0.1$ Gyr ago, where the uncertainties quoted are statistical only. The substructures GL-4 and GR-1, with smaller pericenters, exhibit higher significance in the less massive potential of the Milky Way, implying that the more massive potential may overestimate the central mass, especially the bulge. The accretion times of GL-4 and GR-1 are $6.9 \pm 0.3$ Gyr with a confidence of $3.7\sigma$, and $2.0 \pm 0.1$ Gyr with a confidence of $4.4\sigma$, respectively. Further constraints on the accretion times of phase-space substructures require more precise astrometric data, e.g., by Gaia DR4, China Space Station Survey Telescope and Roman space telescope.
\end{abstract}


\section{introduction}

\par Understanding how the Milky Way assembled its stellar halo is a central objective of Galactic archaeology. In the hierarchical model of galaxy formation, the Milky Way grew through the accretion and subsequent disruption of smaller systems, and the tidal debris typically is found in stellar streams or moving groups \citep{bullock05Tracing,cooper10Galactic}. Numerical simulations predict the solar neighbourhood may contain several hundreds of these streams \citep{helmi99Building,gomez13Streams}. Fossil remnants of past accretion events therefore provide valuable insights into the progenitors' properties and their accretion histories, helping to constrain hierarchical assembly scenarios.

\par The astrometric data provided by Gaia \citep{collaboration16Gaia} have unprecedented number and precision. Its Data Release \citep[][DR2, ]{collaboration18Gaiaa} provided high precision proper motions and parallaxes for over 1.3 billion stars. Gaia DR3 \citep{collaboration21Gaia} increased the sample by $\sim10$\% and improved astrometric precision, especially for proper motions (about a factor of 2.5). When combined with radial velocities from spectroscopic surveys, these data yield full six-dimensional phase-space information that enables identification and characterization of stellar halo substructure.

\par \citet{helmi17Box} revealed a rich and complex structure in integrals-of-motion space, and identified several statistically significant substructures. Thamnos was reported as a low-energy halo structure and is characterized by low-inclination, mildly eccentric retrograde orbits \citep{koppelman19Multiple}. Chemical abundances are commonly used to provide complementary constraints on the origins of the structures. Numerous high-energy and retrograde stellar substructures were identified by \citet{myeong18Discovery}. These were associated, together with several globular clusters, with a substantial and separate accretion event distinct from Gaia-Sausage-Enceladus (GSE), referred to as Sequoia \citep{myeong19Evidence}.

\par Numerical simulations can be used to further probe the progenitors' properties. GSE was identified as a distinct structure by \citet{belokurov18Coformation} and \citet{helmi18Merger}, the latter interpreting it as debris from a major early merger, with a mass ratio of approximately $4:1$. \citet{koppelman19Characterization} revisited the Helmi streams and identified seven globular clusters that are likely associated with them. They used simulations to investigate the progenitor's mass and to constrain its accretion time. \citet{webb19Searching} used simulations to predict two possible accretion histories for the progenitor of GD-1.

\par \citet{gomez10Identificationa} provided an alternative method to explore accretion histories. They used orbital frequencies to estimate the accretion times of satellite debris and validated the method on simulated data. The method is tested in time evolving Plummer potentials and in a fully self-consistent N-body simulation. In both cases the technique reliably recovered disruption times, demonstrating its robustness. \citet{gomez10Identification} further tested the method in N-body simulation that include the apparent magnitude limit, background field star contamination and observational errors, demonstrating the method's feasibility under realistic observing conditions and highlighting the critical role of data precision.

\par In this paper, we use the method of \citet{gomez10Identificationa} to estimate accretion times for six halo substructures reported in previous studies. We introduce a statistical test to assess the significance of our accretion time estimates. The analysis is repeated in two Milky Way potentials to check for systematic sensitivity.

\par This paper is organized as follows. Section~\ref{sec:data} describes the data sets and sample selection. Section~\ref{sec:omega} presents a Bayesian approach for evaluating uncertainties and the procedure for computing orbital frequencies. In Section~\ref{sec:time} we introduce the estimation of accretion times and the evaluation of their uncertainties and significances, and compares the results obtained under two Milky Way potentials. The conclusions are summarized in Section~\ref{sec:conclusions}.

\section{data}
\label{sec:data}

\par In order to estimate the accretion time of a substructure, we need to obtain its member stars. We directly use the six substructures discovered by \citet{li19Substructures} and \citet{li20Two} as our sample. These substructures were identified as overdensities in the space of integrals of motion, defined as the energy and two components of the angular momentum, using Gaia \citep{collaboration16Gaia}, LAMOST \citep[Large Sky Area Multi-object Fiber Spectroscopic Telescope, ][]{zhao12LAMOST,cui12Large} and RAVE \citep[Radial Velocity Experiment, ][]{kunder17Radial} data. In addition to the member stars of the substructures, other halo stars from the above paper were also collected as a background sample for comparison.

\subsection{Improving data precision}

\par We updated the astrometric information from Gaia Data Release 2 \citep[DR2, ][]{collaboration18Gaiaa} in the reference data with that from Gaia Data Release 3 \citep[DR3, ][]{gaiacollaboration23Gaia}. The median parallax (proper motion) uncertainties are $0.02 - 0.03$ mas ($0.02 - 0.03$ mas/yr) for sources with $G<15$, 0.07 mas (0.07 mas/yr) at $G=17$, 0.5 mas (0.5 mas/yr) at $G = 20$, and 1.3 mas (1.4 mas/yr) at $G = 21$ mag \citep{collaboration21Gaia}. Compared with Gaia DR2, parallax precision is improved by about 30 percent, while proper motion precision is improved by a factor of 2. Sources with bad astrometric solutions are excluded using the criterion \texttt{fidelity\_v2}$<0.5$ \citep{rybizki22Classifier}. We further apply the parallax zero-point correction of \citet{lindegren21Gaia}, which is determined based on quasars.

\par Li et al. (2025, in preparation) collected radial velocities from Gaia DR3 and five spectroscopic surveys, and corrected for uncertainties as well as systematic offsets between the surveys. By applying a $\chi^2$ test for the constancy of multiple measurements of the same star, they removed radial velocity variable sources. Combined radial velocities and uncertainties were derived from the weighted average of multiple measurements. Using the catalog they provided, the precision of the radial velocities is improved by a factor of 3.

\subsection{Updating member star lists}

\begin{figure*}[tb!]
	\plotone{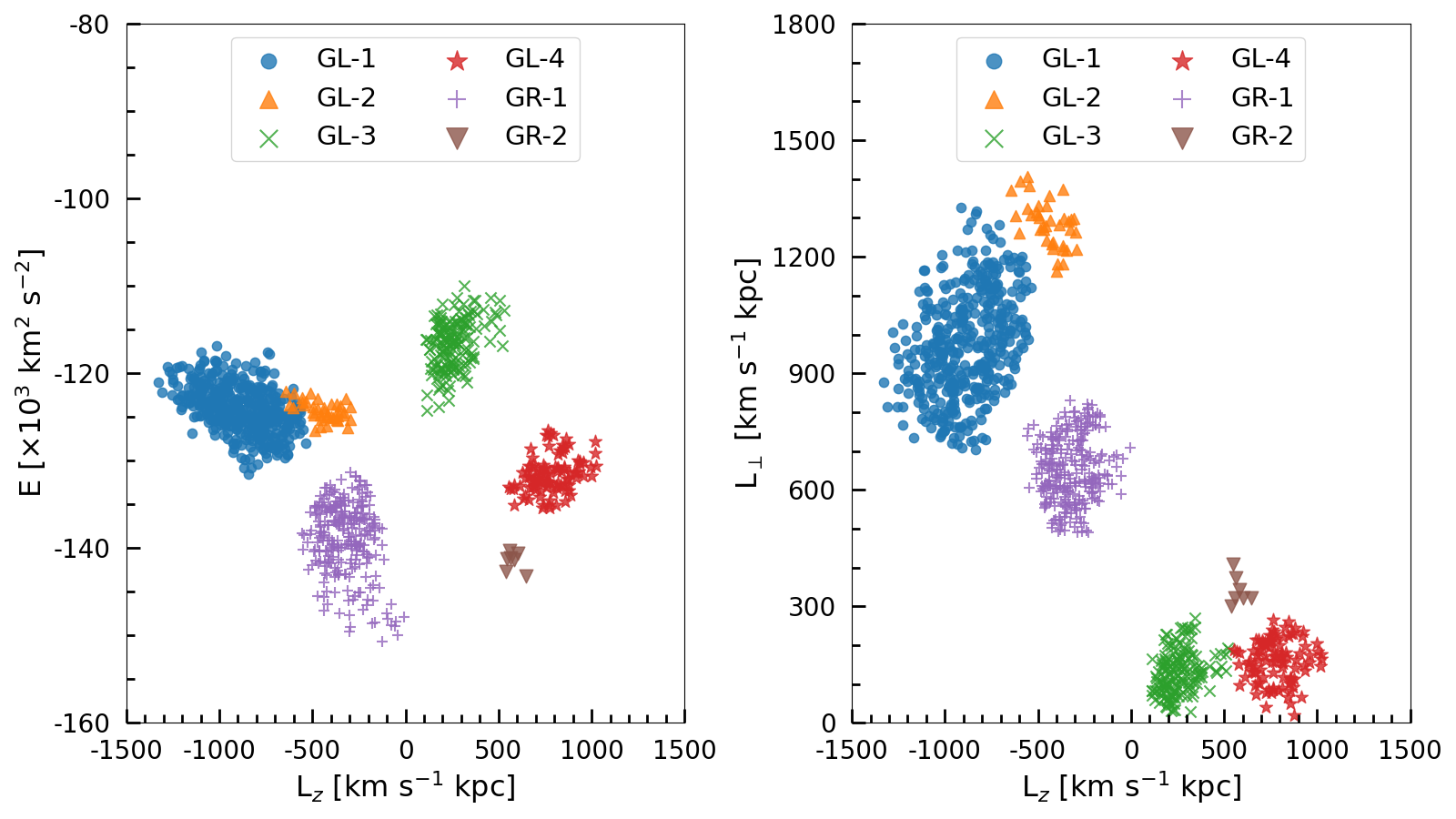}
	\caption{Distribution of member stars of the six substructures in energy $E$ vs. angular momentum $L_z$ space (left panel) and in $L_\perp = \sqrt{L_x^2 + L_y^2}$ vs. $L_z$ space (right panel). Different substructures are shown with distinct colors and markers.}
	\label{fig:ell}
\end{figure*}

\par The Galactocentric Cartesian positions and velocities of each star are recomputed in a right-handed coordinate frame. The $x$-axis points toward the Sun and the $z$-axis points toward the north Galactic pole. In this coordinate system, the Sun is located at $(x_\odot, y_\odot, z_\odot) = (8.277, 0, 0.0208)$ kpc \citep{gravitycollaboration22Mass,bennett19Vertical}, with a velocity of $(v_{\odot,x}, v_{\odot,y}, v_{\odot,z}) = (11.1, 251.5, 8.6)$ km/s \citep{schonrich10Local,reid20Proper}. The velocities in the $y$ and $z$ directions are derived from the proper motion of Sgr A$^*$, assuming it has the same velocity as the Galactic center.

\par We compute the energy $E$ and angular momentum components ($L_z$ and $L_\perp = \sqrt{L_x^2 + L_y^2}$), and re-scale them following the studies that reported these substructures, so that they are on comparable scales. We adopt the potential model derived by \citet{eilers19Circular}, which consists of a Navarro-Frenk-White \citep{navarro97Universal} dark halo, Miyamoto-Nagai \citep{miyamoto75Threedimensional} thin and thick disks, and a spherical Plummer \citep{plummer11Problem} bulge.

\par Outliers are first removed from the original member star lists based on the 3D distances of the substructure members in the scaled space of integrals of motion. For each substructure, we treat every member star as a node and consider two stars to be connected if their 3D distance in the scaled integral-of-motion space falls below a specified threshold. The threshold is taken to be the bandwidth used for the kernel density estimation (KDE) in the original study that identified the substructure. We then retain the largest connected component (the maximal subset in which every star is reachable from every other) as the cleaned member list. Furthermore, six stars originally classified as GL-2 are found to lie closer to the GL-1 region and are therefore manually reassigned to GL-1.

\par New member stars are then identified by computing the 3D distances in the scaled integral-of-motion space between every halo star in the original study and each star in the cleaned member list. Halo stars are identified as new members if they lie within the KDE bandwidth of at least five member stars, with at least one of those lying within half the bandwidth. Figure~\ref{fig:ell} shows the distribution of updated member stars of the six substructures in the space of integrals of motion, with different substructures indicated by distinct colors and markers.

\section{Orbital frequencies and uncertainties}
\label{sec:omega}

\subsection{Bayesian approach}

\par To estimate the uncertainties, a Bayesian approach is applied, with the posterior probability given by:
\begin{align}
P(\bm{\theta}\,|\,\bm{x}) \propto &\exp [-\frac{1}{2} (\bm{x} - \bm{\mu(\theta)})^\mathrm{T} \bm{\Sigma}^{-1} (\bm{x} - \bm{\mu(\theta)})]\  \notag \\ &\times P(d | \alpha , \beta , L) P(v_{\alpha}) P(v_{\delta}),
\end{align}
where $\bm{\theta} = (d,\ v_{\alpha},\ v_{\delta})^\mathrm{T}$, $\bm{x} = (\varpi,\ \mu_{\alpha^*},\ \mu_{\delta})^\mathrm{T}$, $\bm{m} = (1/d,\ v_{\alpha}/kd, \ v_{\delta}/kd)^\mathrm{T}$, $k = 4.74047$ and $\Sigma$ is covariance matrix. The distance prior $P(d | \alpha , \beta , L)$ follows a three-parameter generalized gamma distribution, whose parameters are fitted using the GeDR3 mock catalog. The tangential velocity priors $P(v_{\alpha})$ and $P(v_{\delta})$ are adopted as uniform distributions.

\par For each star, 1000 realizations are drawn from the posterior probability using the Markov chain Monte Carlo sampler \texttt{emcee} \citep{foreman-mackey13Emcee}, employing 16 walkers, running 2770 steps per chain, thinning by a factor of 40, and discarding the first 250 steps as burn-in. The radial velocity is drawn directly from the normal distribution. Each realization is combined with the sky coordinates to yield Galactocentric Cartesian positions and velocities. The ensemble of realizations is used to estimate uncertainties.

\subsection{Orbital frequencies}

\par To compute the orbital frequencies, we first perform orbit integration with \texttt{galpy} \citep{bovy15Galpy} using the \citet{eilers19Circular} potential, integrating each orbit for at least 100 orbital periods. Orbits are integrated using the adaptive time step, and the resulting orbits are sampled at 1 Myr intervals to obtain positions and velocities in the Galactocentric spherical system.

\par The implementation \texttt{naif} \citep{beraldoesilva23Orbital}, based on the Numerical Analysis of Fundamental Frequencies algorithm, is used to compute the orbital frequencies ($\Omega_r, \Omega_\phi$) in the radial ($r$) and azimuthal ($\phi$) directions, respectively, of the Galactocentric spherical system. In this study, we only consider the magnitudes of the frequencies. For realizations with orbital periods longer than 10 Gyr or unbound orbits, the orbital frequencies are set to 0.

\par For the $r$ direction, the real time series $f_r = r$ is used as input for the frequency analysis. For the $\phi$ direction, the complex time series $f_\phi = \cos(\phi) + i\sin(\phi)$, which performs better, is adopted. When computing the $\phi$ direction orbital frequency $\Omega_\phi$, we restrict $\Omega_\phi$ to the range $\Omega_r/2 < \Omega_\phi < \Omega_r$, with the boundaries corresponding to the two limiting cases of a homogeneous sphere and a point mass \citep{gomez10Identificationa}.

\begin{figure}[htb!]
	\plotone{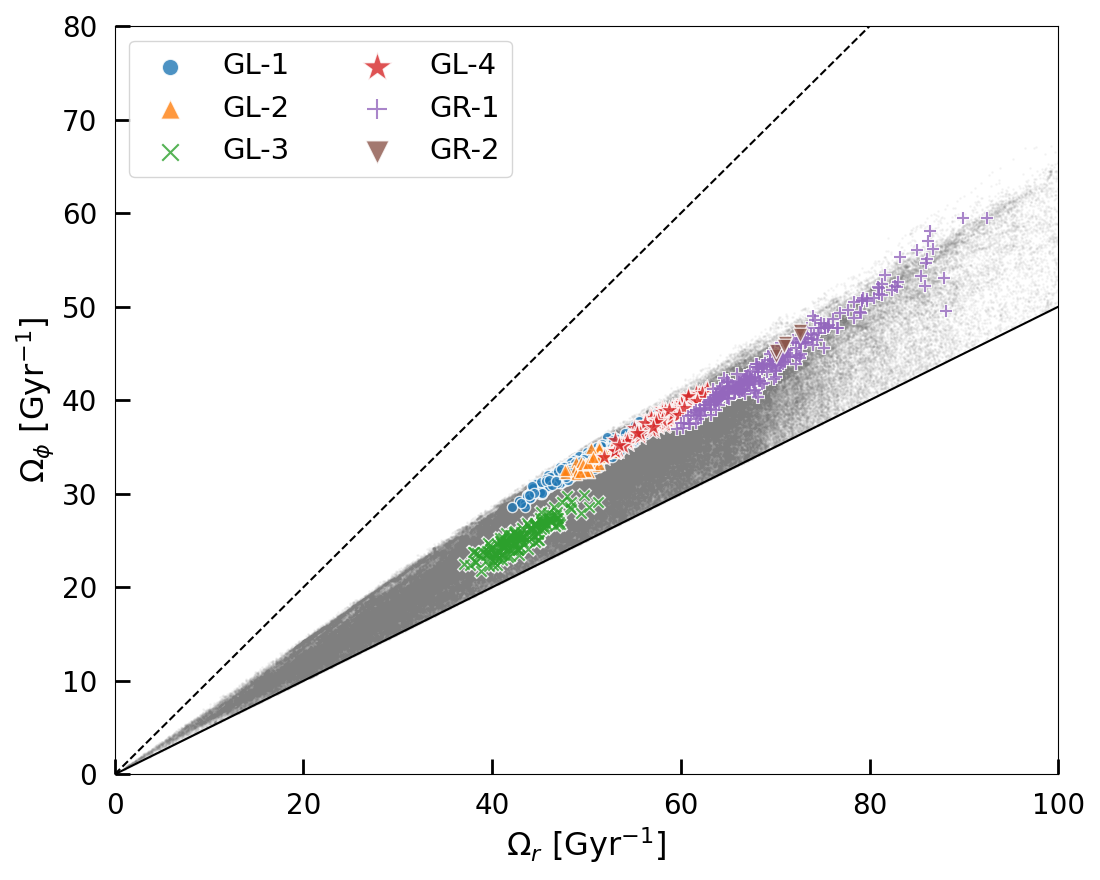}
	\caption{Distributions of halo stars (gray dots) and substructure member stars (colored points) in the orbital frequency space, with colors and markers indicating the different substructures. The dashed and solid lines represent the two limiting cases of a point mass ($\Omega_\phi = \Omega_r$) and a homogeneous sphere ($\Omega_\phi = \Omega_r/2$), respectively.}
	\label{fig:omega}
\end{figure}

\par For each star, we take the median of its 1000 realizations, and estimate the uncertainty as half the difference between the 84th and 16th percentiles. The orbital frequency distributions of the member stars in the six substructures are shown in Figure~\ref{fig:omega}, where different colors and markers represent different substructures. For comparison, we also show in the figure the distributions of the realizations of the other halo stars, indicated by gray dots. Note that, to illustrate the coverage of halo stars in the orbital frequency space, each gray dot corresponds to a single realization, whereas each point in other colors and markers corresponds to a single star.

\begin{figure}[htb!]
	\plotone{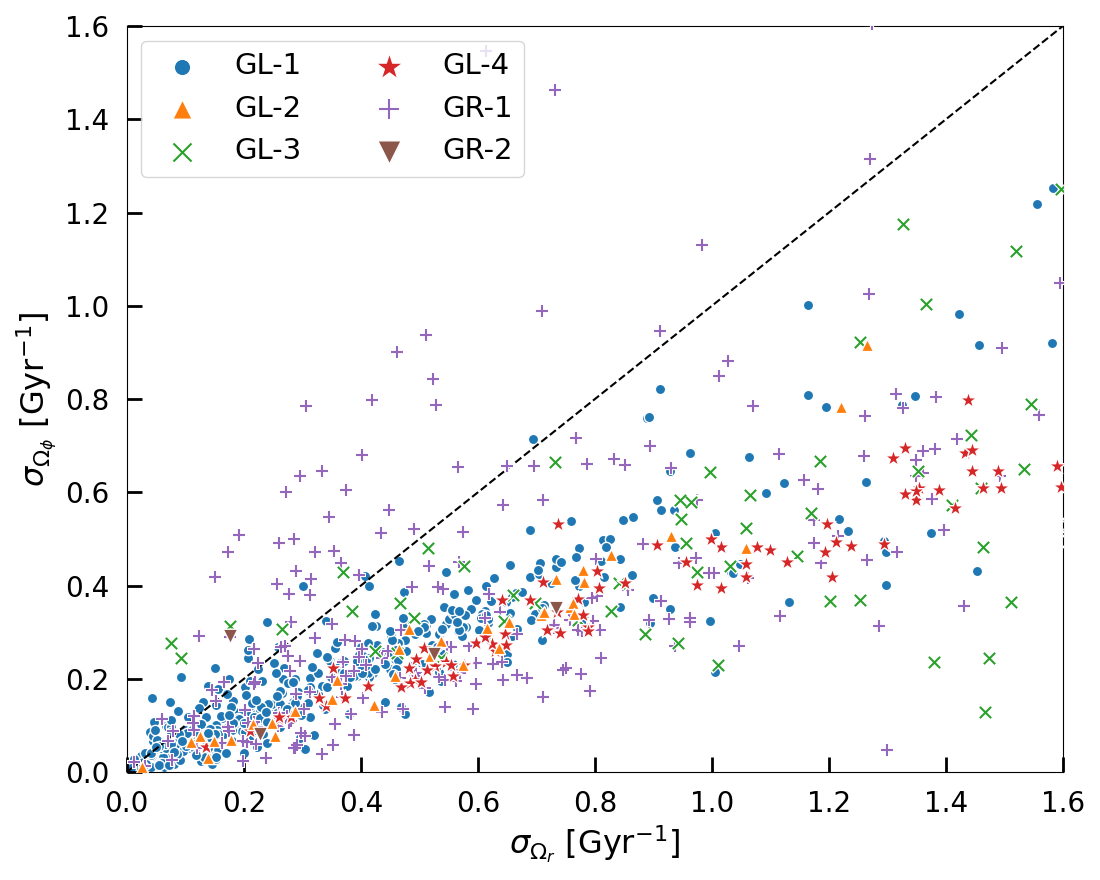}
	\caption{Distributions of the uncertainties in $r$- and $\phi$-direction orbital frequencies for the member stars of the six substructures, with each substructure shown in a different color and marker. The dashed line indicates the 1:1 relation.}
	\label{fig:omega_err}
\end{figure}

\par Figure~\ref{fig:omega_err} shows the scatter of the uncertainties in $r$- and $\phi$-direction orbital frequencies ($\sigma_{\Omega_r}, \sigma_{\Omega_\phi}$) for the member stars of the substructures, with different substructures indicated by different colors and markers. We can see that the precision in $\Omega_\phi$ is generally better than in $\Omega_r$. The substructure GL-1 contains the largest number of member stars and is more concentrated at small uncertainties. GR-1 also has a relatively large number of member stars, but their uncertainties show a broader distribution. GL-2 and GL-4 and have slightly fewer member stars and larger uncertainties. GL-3 exhibits poor accuracy, while GR-2 contains only a small number of member stars, making it difficult to obtain reliable results for either of them.

\section{accretion time}
\label{sec:time}

\subsection{Estimating the time of accretion}

\par \citet{gomez10Identificationa} and \citet{gomez10Identification} proposed a method for estimating accretion times $t_\mathrm{acc}$ from orbital frequencies $\Omega$ and validated it using simulated data. In their definition, the accretion time corresponds to the time when 80 percent of the member stars became unbound from their progenitor. The method exploits that, at a particular spatial location (e.g. the solar neighbourhood), tidal debris appears as multiple clusters in orbital frequency space, and uses the decrease over time in the spacing between adjacent clusters to estimate the accretion time. These clusters result from choosing a particular spatial location, which effectively selects debris with angles $\theta \approx \theta_0 + 2\pi n$ (integer $n$), i.e., the periodic set of angles centered on $\theta_0$. Since
\begin{equation}
\theta (t_\mathrm{acc}) = \Omega t_\mathrm{acc} + \theta (0),
\end{equation}
this also selects a discrete set of particular orbital frequencies $\Omega$, which manifests as multiple clusters in orbital frequency space. For two adjacent clusters, we have
\begin{equation}
\Delta \theta (t_\mathrm{acc}) = \Delta \Omega t_\mathrm{acc} + \Delta \theta(0),
\end{equation}
where $\Delta \theta (t_\mathrm{acc}) = 2\pi$ (i.e. one angular period). If the accretion time $t_\mathrm{acc}$ is sufficiently large, the initial separation $\Delta \theta(0)$ can be neglected, and hence the accretion time is obtained from the orbital frequency difference $\Delta \Omega$ between adjacent clusters:
\begin{equation}
t_\mathrm{acc} \approx \frac{2\pi}{\Delta \Omega}.
\end{equation}

\par Although they were unable to apply the method to real observations available then due to its stringent precision requirements, the high-precision data from Gaia DR3 now provide the necessary foundation to do so. We adjust some details of the method and describe the full procedure below.

\par In step 1, we compute the bi-dimensional histogram in the $\Omega_r - \Omega_\phi$ plane individually for each substructure, with a bin width of $\Delta$ in both dimensions. The bin width $\Delta$ may differ among substructures and will be described later. For both dimensions, the number of bins $N$ is required to be equal, odd, and $\geqslant 201$. To reduce sensitivity to the choice of bin edge positions, we compute histograms for each dimension with the bin edges shifted by 0, 0.2, 0.4, 0.6, and 0.8 times the bin width $\Delta$, and constructed all combinations of these shifts across the two dimensions.

\begin{figure*}[htb!]
	\plottwo{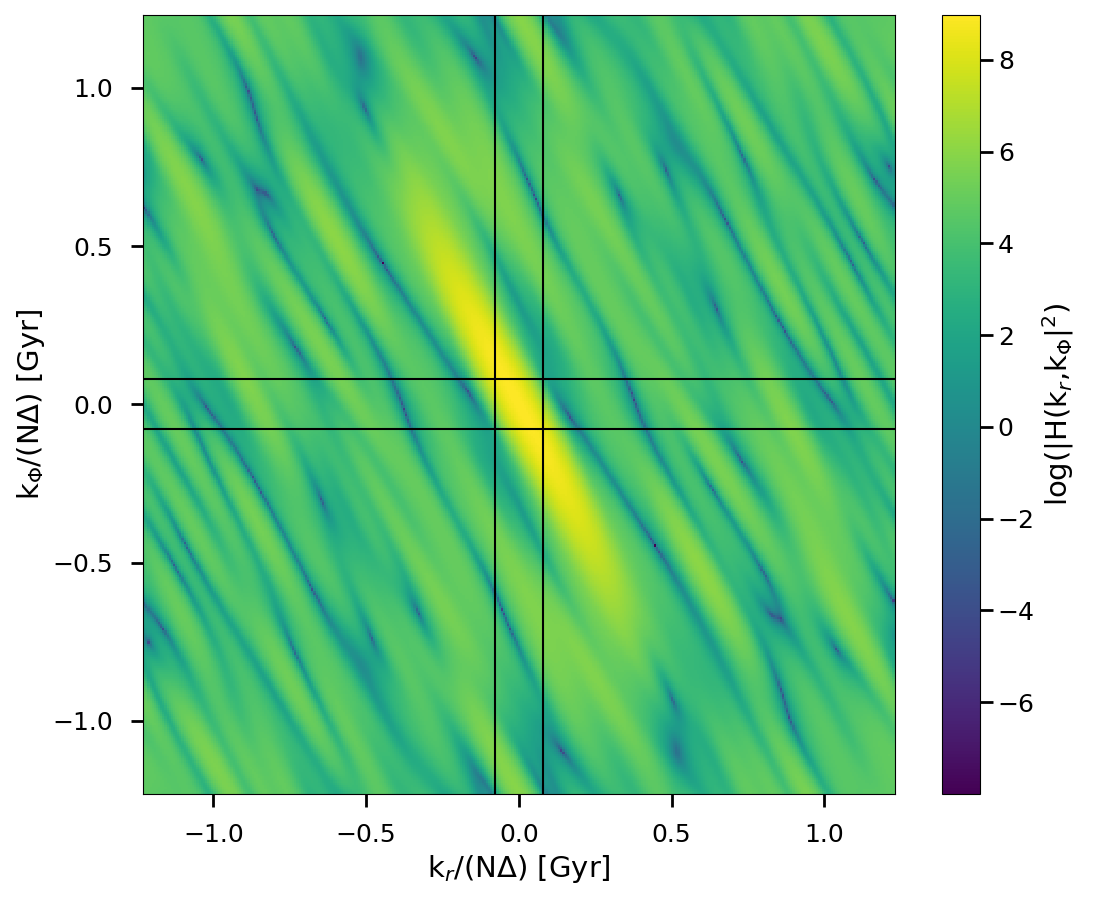}{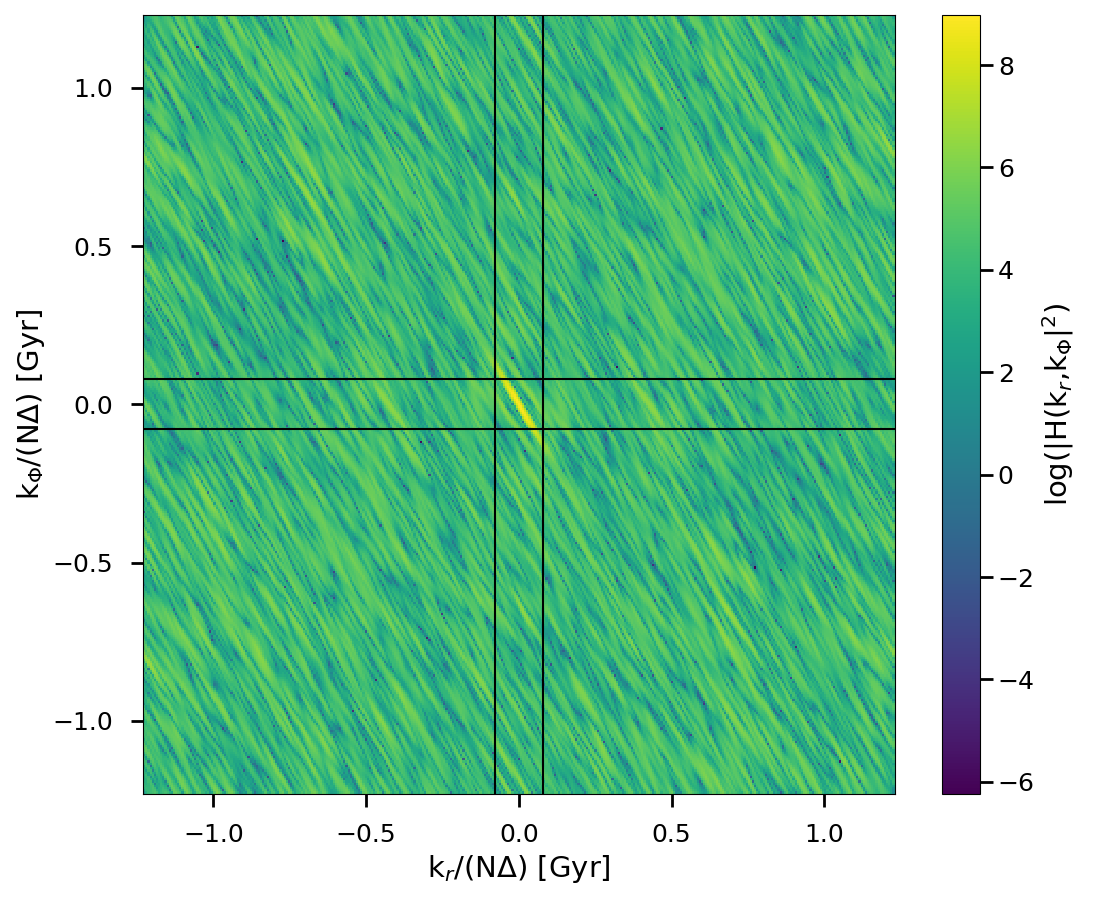}
	\caption{Power spectra of the Fourier transform for substructure GL-1 (left panel) and a random sample (right panel). The axes represent the wavenumbers in the two direction. The black lines mark the regions used to compute the 1-D power spectra.}
	\label{fig:h2}
\end{figure*}

\par In step 2, we compute the 2-D discrete Fourier Transform of the bi-dimensional histogram, denoted as $H(k_r, k_\phi)$. As an example, the power spectrum $|H(k_r, k_\phi)|^2$ of substructure GL-1, which is color coded, is shown in the left panel of Figure~\ref{fig:h2}. The coordinates indicate the wavenumbers in each direction $(k_r, k_\phi)$. For comparison, the right panel shows the power spectrum of a random sample constructed by drawing from the halo the same number of stars as in GL-1. It can be seen that, compared with those in the right panel, the bright streaks in the left panel are wider and longer. While the feature is most pronounced in the center, it is also visible along the edges of the figures.

\par In step 3, we marginalize the 2-D power spectrum $|H(k_r,k_\phi)|^2$ over each axis within a narrow band around 0, and combine positive and negative wavenumbers to obtain the one-dimensional power spectrum along each direction. Specifically,
\begin{equation}
P(k_r)=
\begin{cases}
	\dfrac{1}{N^2} \displaystyle\sum_{k_\phi=-k_\mathrm{slit}}^{k_\mathrm{slit}} |H(0,k_\phi)|^2, \\
	\hfill \mathrm{if}\; k_r=0,\\[0.6em]
	
	\dfrac{1}{N^2} \displaystyle\sum_{k_\phi=-k_\mathrm{slit}}^{k_\mathrm{slit}} \left( |H(k_r,k_\phi)|^2 + |H(-k_r,k_\phi)|^2 \right),\\
	\hfill \mathrm{if}\; k_r>0,
\end{cases}
\end{equation}
where $k_\mathrm{slit}$ corresponds to the half-widths of the marginalization windows (slits), which are delimited by the black lines in Figure~\ref{fig:h2}. The 1-D power spectrum $P(k_\phi)$ is defined analogously. We fix the slit width to 0.16 Gyr in both directions, so the bin half-width is given by $k_\mathrm{slit} = \lfloor 0.08 N \Delta \rfloor$, where $\lfloor x \rfloor$ denotes the floor function. The characteristic wavenumber $k_*/(N\Delta)$, at which 1-D power spectrum attains its maximum, provides an estimate of the accretion time through $t_\mathrm{acc}=2 \pi k_*/(N\Delta)$.

\subsection{Significances and uncertainties}

\par To estimate the significances and uncertainties of the accretion time results, the above method is applied to the 1,000 realizations of both the substructures and the random samples. For the substructures, we only consider member stars whose orbital frequencies have small uncertainties, with thresholds varying among substructures as specified below.

\begin{figure*}[htb!]
	\plottwo{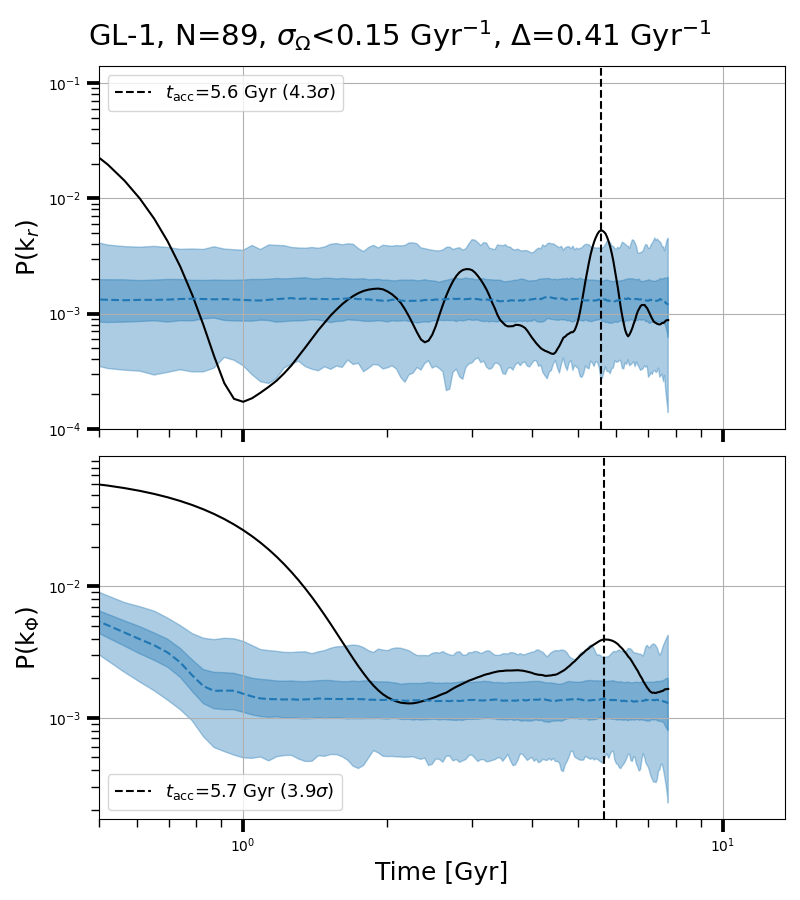}{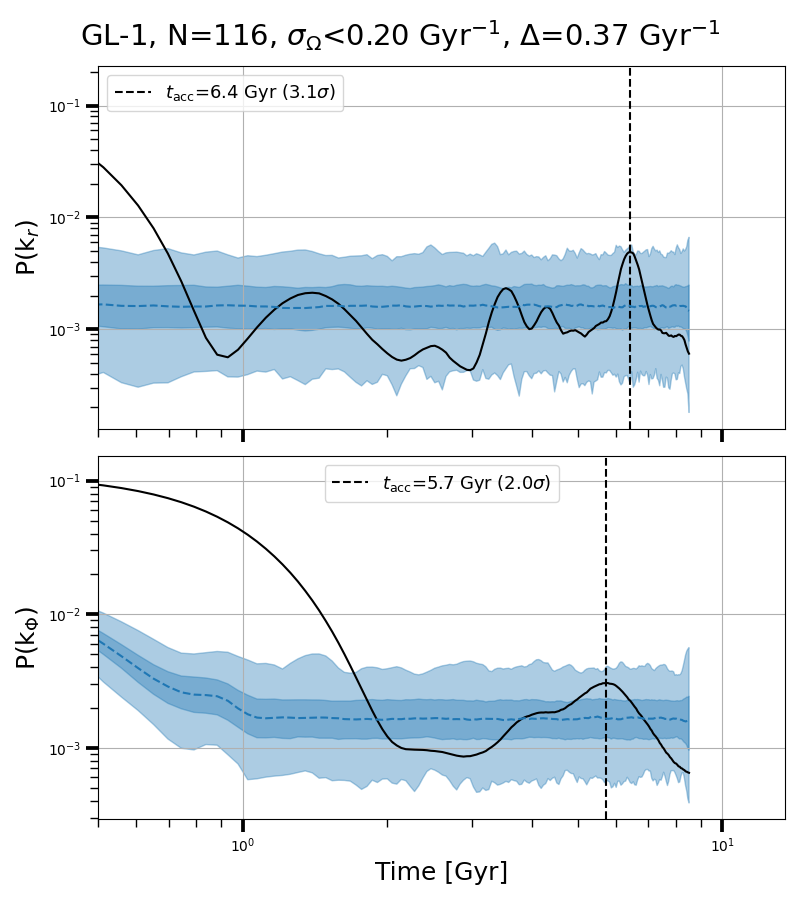}
	\caption{Median 1-D power spectra for substructure GL-1 (black solid curves) and random samples (blue dashed curves). The $1\sigma$ and $3\sigma$ intervals of the random samples are shown as dark blue and light blue shaded regions, respectively. The $x$-axes represent $2\pi$ times the wavenumbers in the $r$ (top panels) and $\phi$ (bottom panels) directions. The estimated accretion times are marked by black vertical dashed lines, with significances given in the legend. Left and right panels show the results using the \citet{eilers19Circular} and \citet{bovy15Galpy} potentials, respectively. Subtitles indicate the substructure name, number of stars, bin width $\Delta$, and applied uncertainty selection.}
	\label{fig:ps_gl1}
\end{figure*}

\par Using the substructure GL-1 as an example, we adopt a bin width $\Delta$ of 0.41 and select member stars with $\sigma_{\Omega_r}<0.15$ Gyr$^{-1}$ and $\sigma_{\Omega_\phi}<0.15$ Gyr$^{-1}$ to estimate the accretion time. The choice of the bin edge shifts will be discussed later. A total of 1,000 sets of 1-D power spectra $P(k_r)$ and $P(k_\phi)$ are computed, but the corresponding accretion times are not calculated at this stage. The black solid curves on the left panels of Figure~\ref{fig:ps_gl1} show the medians of the power spectra as functions of $2\pi$ times the wavenumber. With this choice of the $x$-axis, the accretion time can be obtained directly. The black vertical dashed lines mark the locations of the maximum peaks, corresponding to accretion times of 5.6 Gyr and 5.7 Gyr for the $r$ and $\phi$ directions, respectively.

\par Applying the above selection on orbital frequency uncertainties results in a sample of 89 stars. For comparison, we randomly draw an equal number of stars from the halo sample 1,000 times. We then compute the 1-D power spectra of these samples, and the medians of the spectra are indicated by the blue dashed curves in the left panels of Figure~\ref{fig:ps_gl1}. The dark blue and light blue shaded regions represent the $1\sigma$ and $3\sigma$ intervals, respectively. It can be seen that, at larger wavenumbers, the power spectra are nearly constant, corresponding to contributions from random noise.

\begin{figure*}[htb!]
	\plottwo{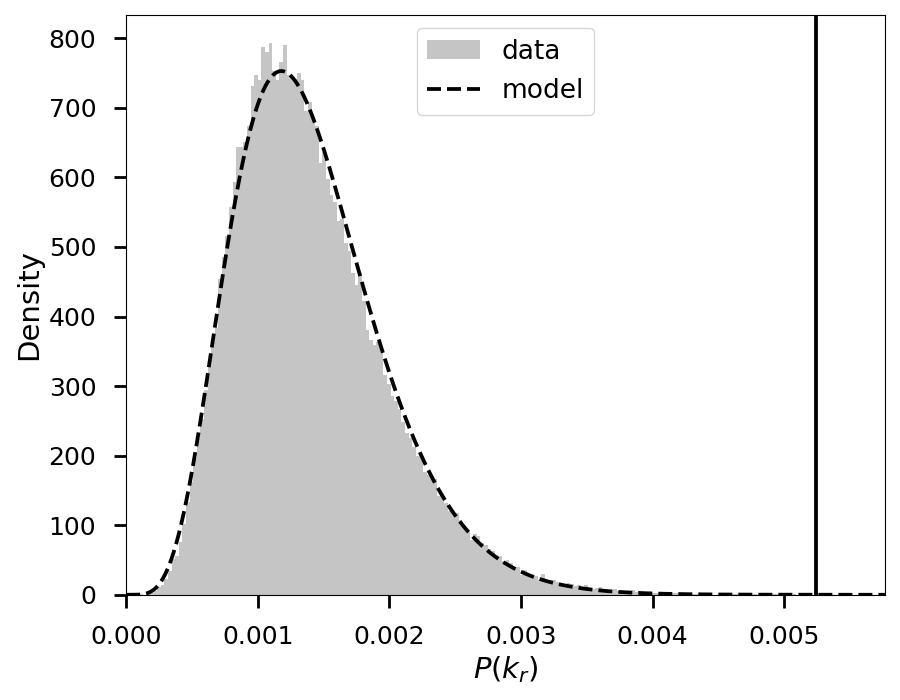}{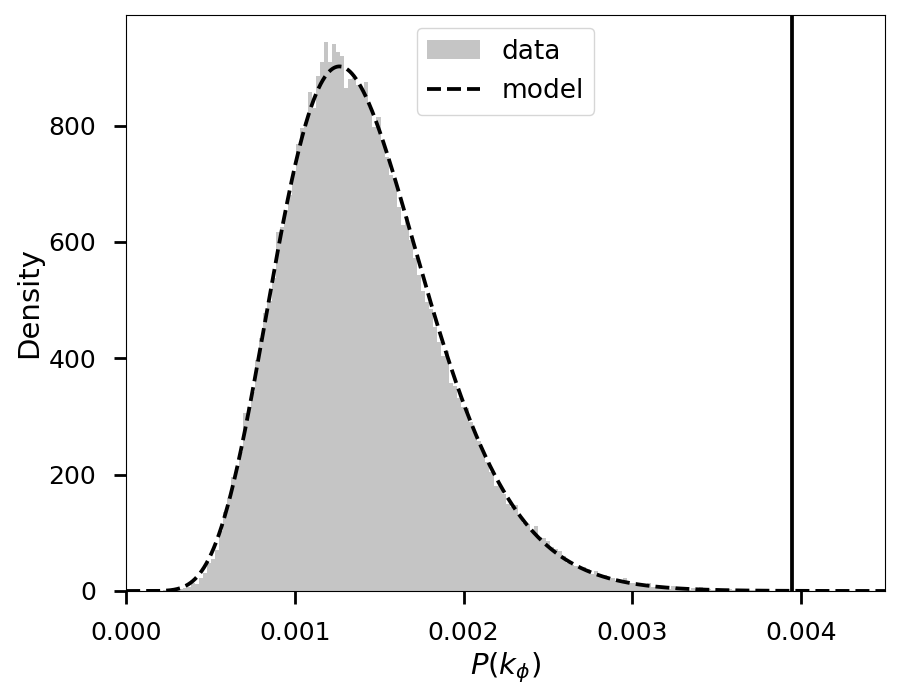}
	\caption{Histograms of the $r$ (left) and $\phi$ (right) direction power spectra in the high wavenumber region. The dashed lines represent the fitted models of the two-parameter gamma distribution. The vertical solid lines mark the maximum peak of the median power spectra of substructure GL-1.}
	\label{fig:gamma}
\end{figure*}

\par The significances of the accretion times inferred for GL-1 can be evaluated by comparing its power spectra with those obtained from the random samples. We fit a two-parameter gamma distribution to the high wavenumber tails of the 1-D power spectra from all random samples. The shape and scale parameters are obtained by maximum likelihood estimation and are fitted separately for the $r$ and $\phi$ directions. We define the lower bound of the high wavenumber tail as the location of the first intersection between the median power spectrum of the substructure and that of the random samples. If no intersection exists, the first local minimum of the median power spectrum of the substructure is taken as a substitute.

\par The fitting results are shown in Figure~\ref{fig:gamma}. The significance of the highest peak in the median power spectrum of the substructure is estimated using the cumulative distribution function, and the peak is indicated by the black vertical line in the figure. In the case of GL-1, the accretion times in the $r$ and $\phi$ directions have significances of $4.3\sigma$ and $3.9\sigma$, respectively.

\par Uncertainties in the accretion time are estimated from the 1,000 sets of 1-D power spectra associated with the realizations. Due to errors, the peak in each realization's 1-D power spectrum that near the highest peak of the median power spectrum is not always the highest peak. Therefore, we adopt a prior that constrains each realization's accretion time estimate to lie around the value derived from the median power spectrum. This prior is implemented as a uniform distribution over a finite interval.

\par To obtain this interval, the median power spectrum of GL-1 is smoothed using \texttt{gaussian\_filter1d} of \texttt{SciPy} \citep{virtanen20SciPy}. We adopt a full width at half maximum of 0.2 Gyr, corresponding to $0.2/2\pi \approx 0.03$ Gyr in wavenumber space. The upper and lower limits of the interval are defined as the nearest troughs on either side of the highest peak, if the median of the power spectra of the troughs is lower than that of the random samples. If no trough on the left side of the peak satisfies the condition, the minimum on that side is taken as a substitute. In the analogous case on the right side, the upper bound is taken to be infinite.

\par For each individual power spectrum, the accretion time is determined within this interval, with the procedure applied separately for the $r$ and $\phi$ directions. The accretion time of GL-1 is estimated to be $5.6 \pm 0.1$ Gyr from the $r$ direction and $5.7_{-0.4}^{+0.2}$ Gyr from the $\phi$ direction. The resulting combined accretion time is $5.6 \pm 0.1$ Gyr. The quoted uncertainties reflect only the contribution from the observational errors in the data.

\par We determine the uncertainty threshold, the bin width $\Delta$, and the bin edge shifts by a 4D grid search to maximize the significance. The uncertainty threshold and $\Delta$ are taken from $e^{-2} \approx 0.14$ to $e \approx 2.72$, with steps of 0.1 dex in log space. The bin edge shifts along each histogram dimension are taken as 0, 0.2, 0.4, 0.6 and 0.8 times the bin width. We only consider results with more than 20 stars after the uncertainty selection. Additional systematic uncertainties may arise from analysis choices such as the bin width and slit size. Moreover, systematic uncertainties introduced by the choice of potential are typically more significant.

\subsection{Effect of different potential}

\begin{figure}[htb!]
	\plotone{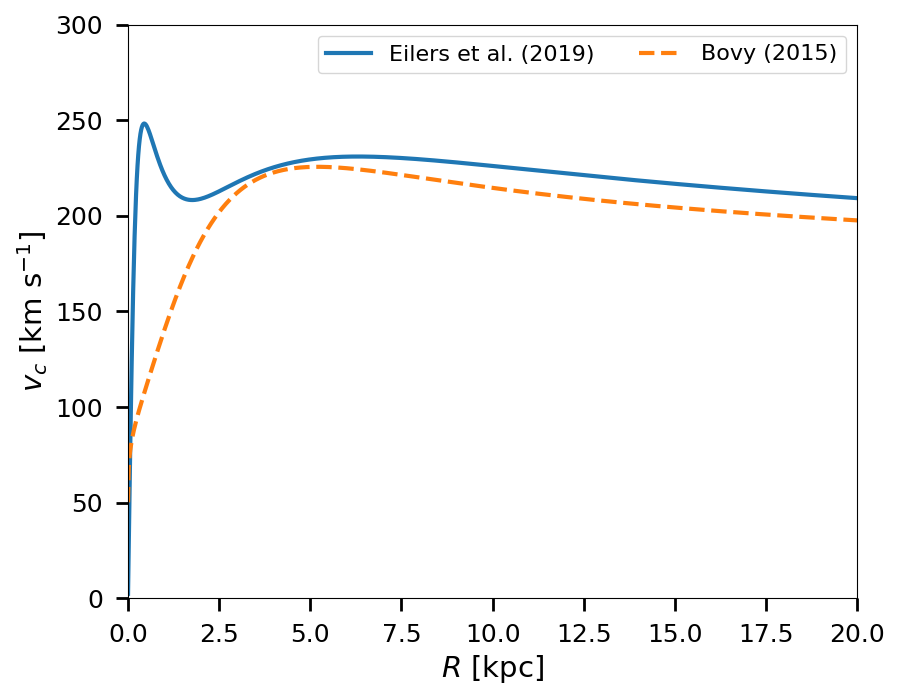}
	\caption{Circular velocity curves $v_c$ of the \citet{eilers19Circular} and \citet{bovy15Galpy} potentials as a function of Galactocentric radius $R$.}
	\label{fig:rc}
\end{figure}

\par The estimated accretion times vary with the choice of the Milky Way potential model. For comparison, we adopt the potential of \citet[][, hereafter B15]{bovy15Galpy}. The circular velocity curves of this potential and the \citet[][, hereafter E19]{eilers19Circular} potential used above are shown in Figure~\ref{fig:rc}. As shown in this figure, the B15 potential has a smaller mass than the E19. At the solar position, the circular velocity is 219 km s$^{-1}$ for the B15 potential and 229 km s$^{-1}$ for the E19 potential.

\par The bin width $\Delta$ and the uncertainty threshold are adjusted to maximize the significance. The right panels of Figure~\ref{fig:ps_gl1} show the 1-D power spectra of GL-1 computed using the B15 potential. It can be seen that the significances are noticeably reduced, especially for the $r$ direction. This reduction occurs when the adopted Milky Way potential poorly represents the true potential \citep{mcmillan08Disassembling}. Nevertheless, the estimated accretion times remain close, suggesting that an incorrect potential is unlikely to fully obscure the true accretion time.

\par Since the E19 potential is based on more reliable Gaia data and the corresponding accretion time estimate exhibits higher significance, we adopt the result derived from the E19 potential.

\par In contrast to the two static potentials used in our analysis, the true potential of the Milky Way is time-dependent. The time-independent approximation is no longer reliable after $2 \sim 3$ Gyr of orbital integration \citep{arora22Stability}. \citet{gomez10Identificationa} found in numerical simulations that accretion times inferred using time-independent approximations are underestimated by roughly 15-25\% relative to the true accretion times, the exact bias depending on the speed of the potential's evolution. This is because, as galaxies grow in mass over time, orbits at a given radius have shorter periods (higher orbital frequencies). Consequently, the spacing between the clumps in orbital frequency space used to estimate the accretion time also increases.

\subsection{Accretion times of other substructures}

\par For substructures GL-2, GL-3 and GR-2, no reliable results were obtained, due to their small numbers of member stars and the low precision of their orbital frequencies.

\begin{figure*}[htb!]
	\plottwo{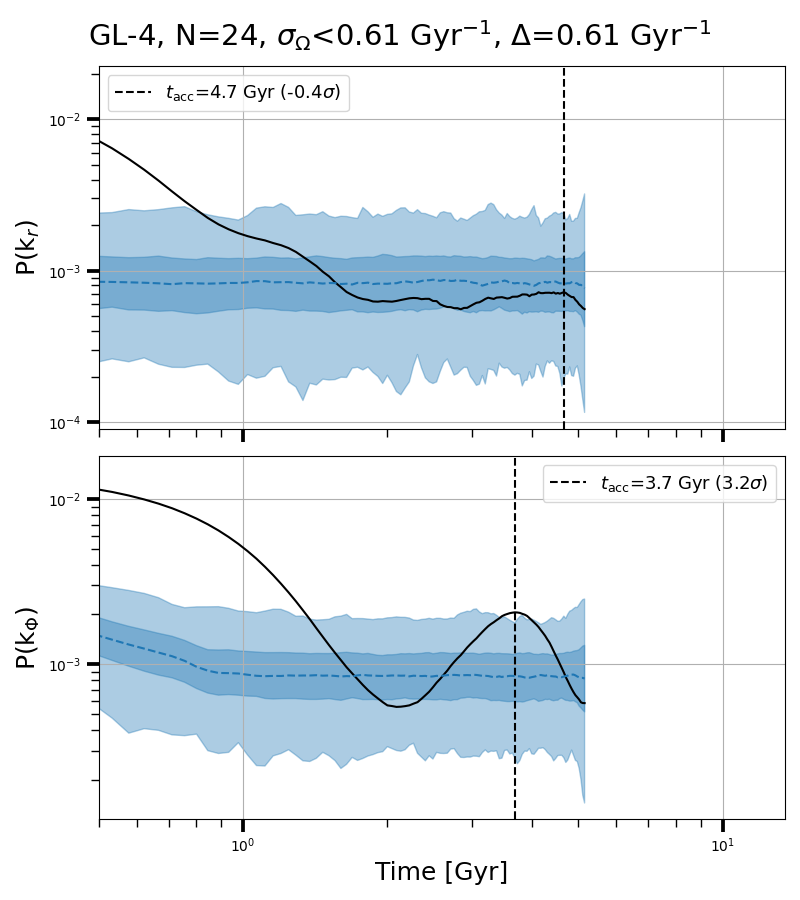}{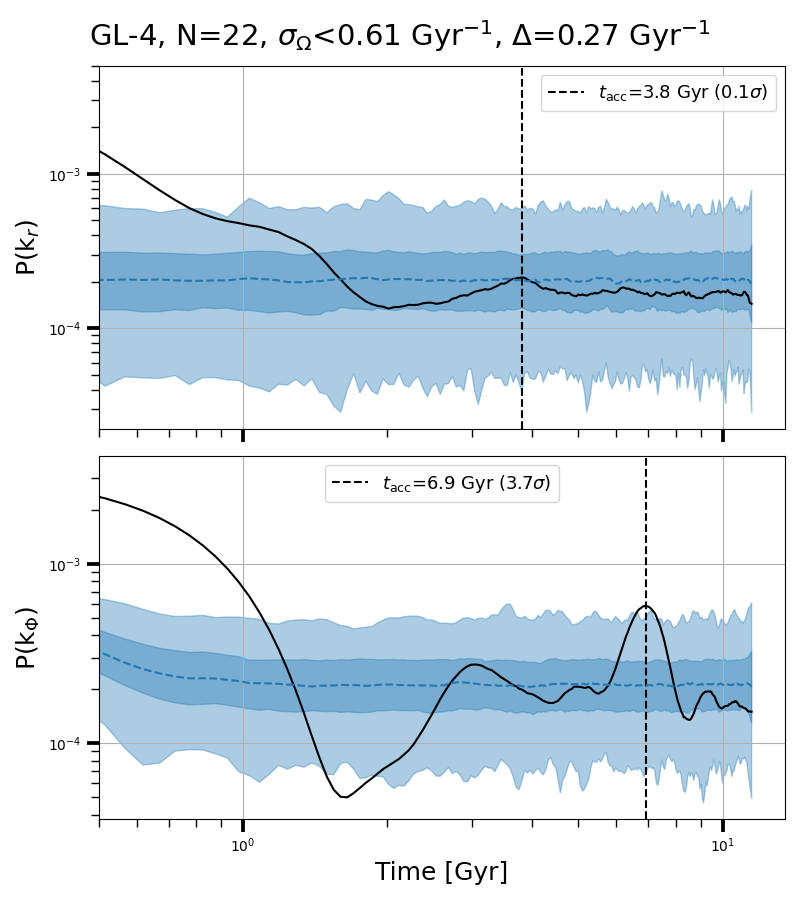}
	\caption{Similar to Figure~\ref{fig:ps_gl1}, but for substructure GL-4.}
	\label{fig:ps_gl4}
\end{figure*}

\par The results for GL-4 are shown in Figure~\ref{fig:ps_gl4}. As shown in the top panels, the significance of the $r$ direction results is very low for both potentials. This is due to the large uncertainties of $\Omega_r$, which make the results to be more similar to those of the random samples. Because the uncertainties of $\Omega_\phi$ are smaller, the $\phi$ direction results exhibit higher significance, even under the same uncertainty threshold.

\par For GL-4, the accretion time inferred from the $\phi$ direction analysis under the B15 potential is $6.9 \pm 0.3$ Gyr, with a significance of $3.7\sigma$. For the E19 potential, the accretion time is estimated to be $3.7 \pm 0.2$ Gyr, with a lower significance of $3.2\sigma$. The accretion time of the latter is approximately half that of the former, whereas the bin width $\Delta$ used in the analysis is approximately twice as large. This may be attributed to larger systematic uncertainties in the orbital frequencies of the E19 potential, which necessitate the use of a bin width twice as large to absorb the impact of these uncertainties. However, increasing the bin width by a factor of two degrades the resolution in orbital frequency space. As a result, the fundamental wavenumber becomes unresolvable, and only the component at half the fundamental wavenumber can be identified. As the accretion time scales linearly with the wavenumber, the inferred accretion time is therefore reduced by a factor of two.


\begin{figure*}[htb!]
	\plotone{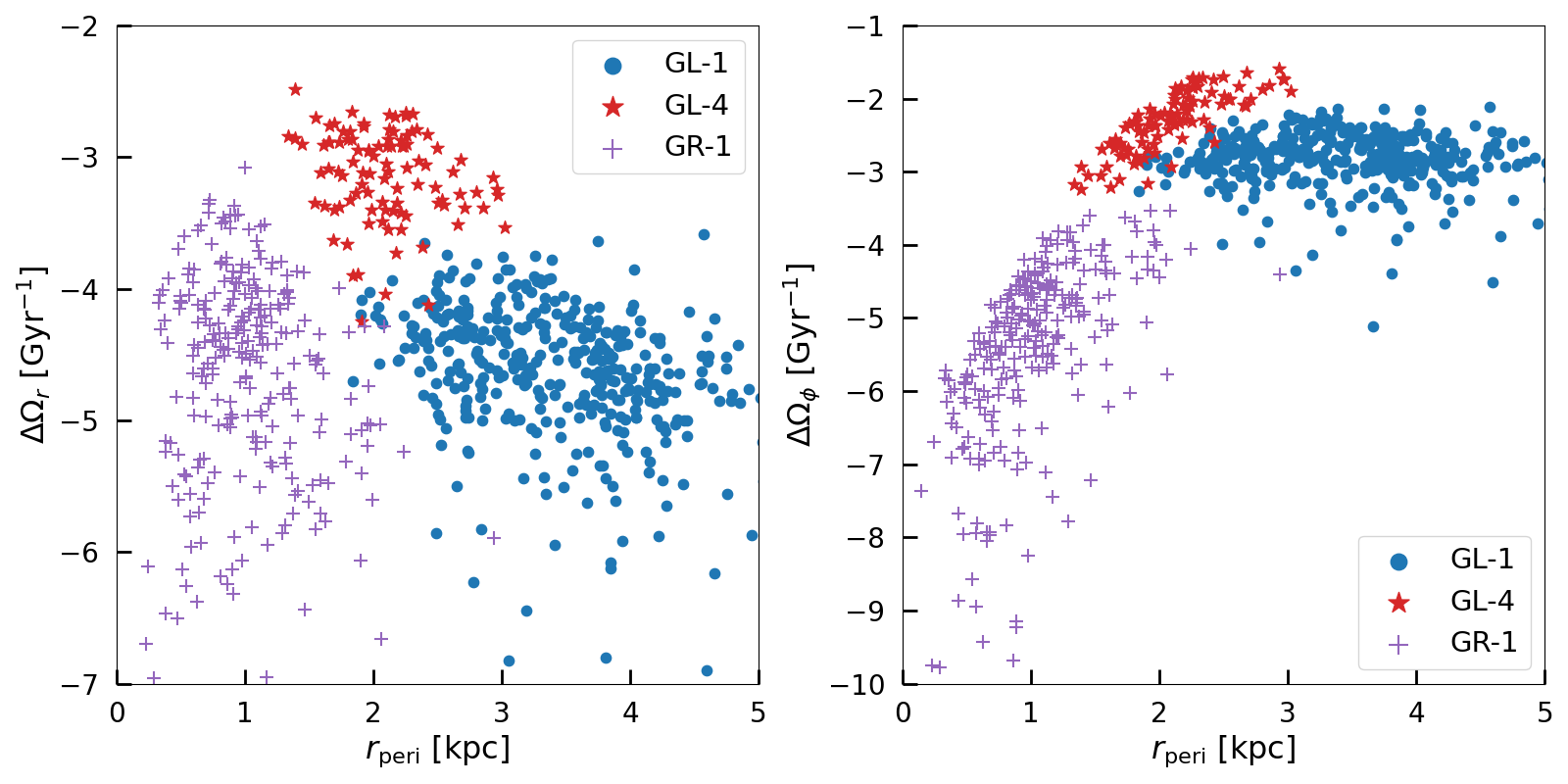}
	\caption{The differences in orbital frequencies between the \citet{eilers19Circular} and the \citet{bovy15Galpy} potentials ($\Omega_\mathrm{Bovy} - \Omega_\mathrm{Eilers}$) for member stars of different substructures versus orbital pericenters under the \citet{eilers19Circular} potential. Different substructures are indicated by different colors and markers. Left and right panels show the frequency differences in the $r$ and $\phi$ directions, respectively.}
	\label{fig:domega}
\end{figure*}

\par The larger systematic uncertainties in the orbital frequencies under the E19 potential likely stem from the smaller pericenters of the GL-4 member stars, where the two potentials differ more. Given that results based on the B15 potential show higher significance, the E19 potential therefore appears to be more strongly offset from the true potential in this region. Figure~\ref{fig:domega} shows the distribution of the differences in orbital frequencies between the two potentials, $\Delta_\Omega = \Omega_\mathrm{Bovy} - \Omega_\mathrm{Eilers}$, versus the orbital pericenters in the E19 potential. Member stars of different substructures are shown in different colors and markers. A negative $\Delta\Omega$ indicates that the orbital frequency is smaller in the B15 potential.

\begin{figure*}[htb!]
	\plottwo{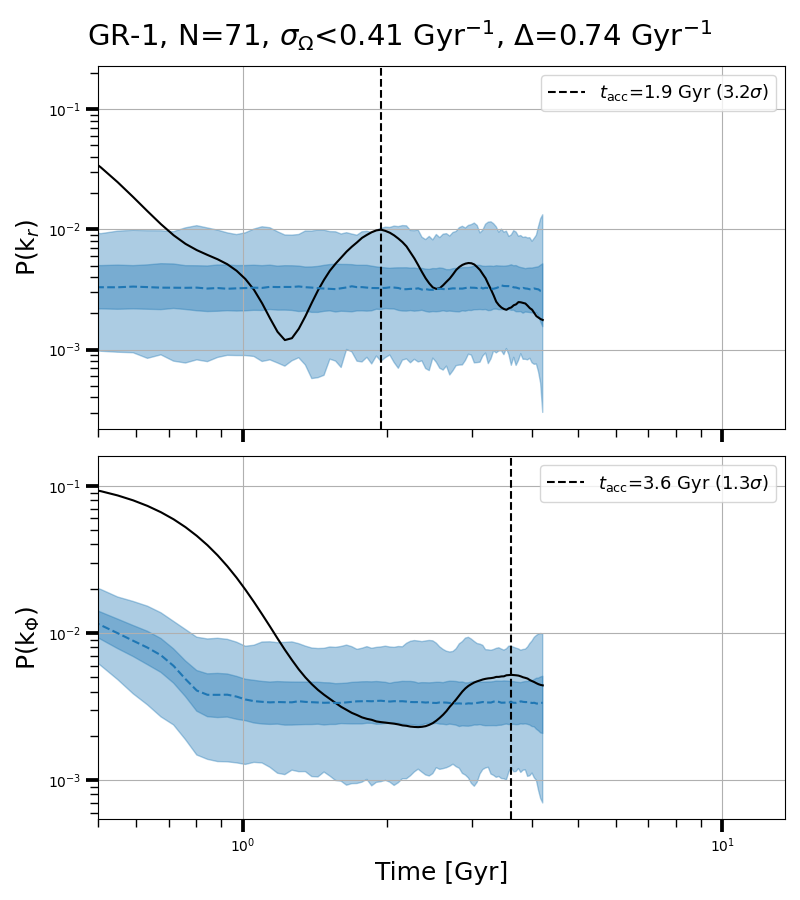}{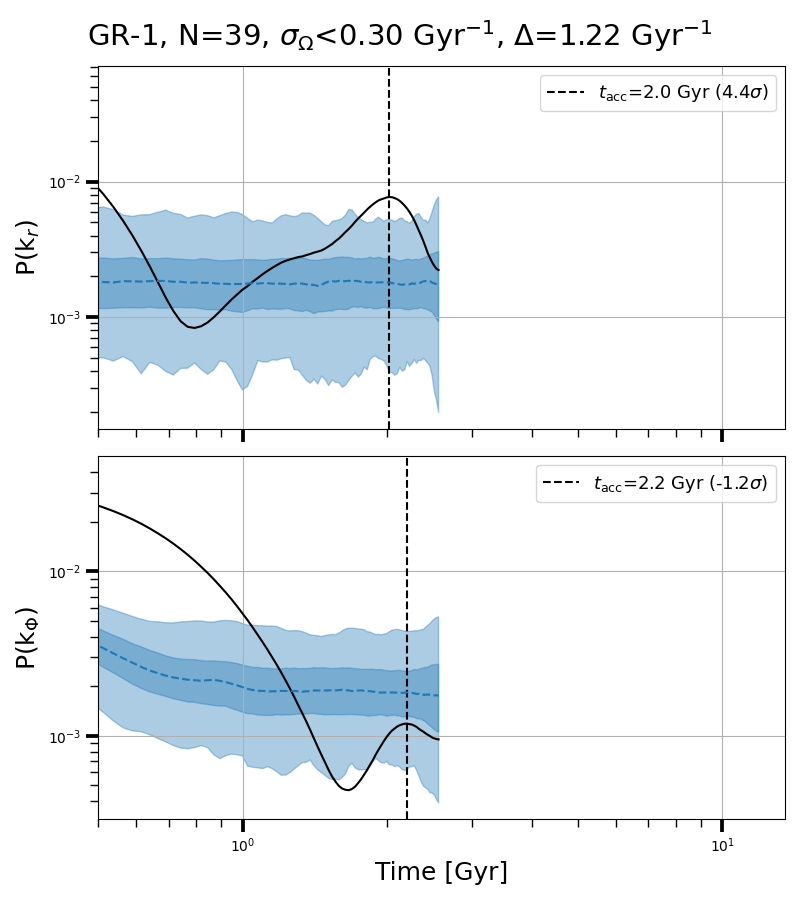}
	\caption{Similar to Figure~\ref{fig:ps_gl1}, but for substructure GR-1.}
	\label{fig:ps_gr1}
\end{figure*}

\par Furthermore, this difference can also provide guidance for improving the potential. \citet{eilers19Circular} adopted the same baryonic components as \citet{pouliasis17Milky} and fitted the dark matter halo using the circular velocity curve within $5 \leqslant R \leqslant 25$ kpc. The region $R<5$ kpc lies outside the fitted domain, and therefore the E19 potential is unreliable there. Since the result based on the B15 potential shows higher significance, the E19 potential may overestimate the mass within $R \sim 5$ kpc. Although the data used to fit the B15 potential also provide poor constraints in the inner region $R<5$ kpc, the B15-based result shows higher significance, suggesting that the E19 potential may overestimate the mass within $R \sim 5$ kpc.

\par However, this does not imply that the B15 potential is reliable near the Milky Way center, since the orbits of GL-4 member stars cover the range 2 kpc $< r < 9$ kpc. Assuming the orbital frequency results obtained in the B15 potential are correct, and that the rotation curve of the B15 potential is underestimated beyond $r=5$ kpc, then the rotation curve inside this radius would require an additional adjustment to offset for that underestimation.

\par A similar situation occurs for GR-1, as shown in Figure~\ref{fig:domega}, the pericenters of its member stars are closer to the Milky Way center compared to those of GL-4, which is also the region where the E19 and B15 potentials exhibit the largest discrepancy. Its 1-D power spectra are shown in Figure~\ref{fig:ps_gr1}. The results from the two potentials are $1.9 \pm 0.1$ and $2.0 \pm 0.1$ Gyr in the $r$ direction, which are fairly close. For the E19 potential, the significance is $3.2\sigma$, whereas for the B15 potential, it is higher, at $4.4\sigma$.

\par For the E19 potential, the power spectrum in the $\phi$ direction shows a trough near 1.9 Gyr. The inconsistency of the E19 potential results between the two directions suggests a limitation of the potential. Under the B15 potential, a small peak appears in the $\phi$ direction power spectrum near 2.0 Gyr, which hints that the E19 potential may overestimate the rotation curve for $R < 5$ kpc. This is consistent with the inference based on the GL-4 results. However, the significance of the results based on the B15 potential is also low, indicating that the $\phi$ direction is more sensitive to the accuracy of the potential model.

\par The pericenters of the member stars of GR-1 are very close to the Milky Way center, where the potential is dominated by the bulge. The comparison between the E19 and B15 potential results may suggest a less massive bulge, i.e., with a mass smaller than the $1.1 \times 10^{10}$ $M_\odot$ of the bulge in the E19 potential. For the B15 potential with a bulge mass of $0.5 \times 10^{10}$ $M_\odot$, the peak around 2.0 Gyr in the $\phi$ direction is also very low, indicating that the bulge is likewise questionable.

\par The most direct way to improve accretion time estimates is to increase data precision, for example with Gaia's upcoming DR4 or future facilities such as the China Space Station Survey Telescope and the Roman Space Telescope. Combining multiple spectroscopic surveys will not only substantially improve the precision of radial velocities but also better identify and remove radial velocity variables (e.g. binaries).

\par Increasing the number of member stars is another way, because when building histograms of orbital frequencies a larger bin width can absorb uncertainties and thus reduce their impact. However, a larger bin width also absorbs the intrinsic stochasticity of random samples and increases power spectra in wavenumber space, so a larger number of member stars is required to recover statistical significance. Moreover, for substructures accreted at earlier times the spacing between clumps in orbital frequency space is smaller, and therefore a larger bin width cannot be applied. Finally, even for GL-1, where the current result is already robust, additional chemical information can further improve the constraints by removing contaminants from the member stars and by directly tightening limits on the accretion time.

\section{conclusions}
\label{sec:conclusions}

\par We have collected six substructures and their member stars from previous studies, and improved their kinematic accuracy by combining Gaia DR3 astrometric data with radial velocities from the stellar catalog compiled by Li et al. (2025, in preparation) from multiple spectroscopic surveys. In the E19 potential, the orbital frequencies of stars in the $r$ and $\phi$ directions were calculated and subsequently used to estimate the accretion time of the substructures. We determine the uncertainties using Monte Carlo method, and establish the statistical significance by comparison with results from random samples drawn from the halo.

\par For the substructure GL-1, the accretion time results based on the orbital frequencies in the $r$ and $\phi$ directions show good consistency, yielding $5.6 \pm 0.1$ Gyr and $5.7_{-0.4}^{+0.2}$ Gyr respectively. The statistical significance is also high, at $4.3\sigma$ and $3.9\sigma$, respectively. The agreement between the two directions and their high significance indicate the robustness of the results. The accretion time for GL-1, obtained by combining the $r$ and $\phi$ direction estimates, is $5.6 \pm 0.1$ Gyr. When the B15 potential is used instead of the E19 potential, which is based on the more reliable Gaia data, the significances decrease substantially, indicating that the results are sensitive to the choice of potential.

\par For GL-4 the inferred accretion time is $6.9 \pm 0.3$ Gyr with a significance of $3.7\sigma$, while for GR-1 it is $2.0 \pm 0.1$ Gyr with a significance of $4.4\sigma$. High significance is observed in only a single direction, and the significance is greater under the B15 potential. This may be because their member stars have pericenters closer to the Milky Way center, where the E19 potential is extrapolated beyond its fitted range and thus overestimates the mass at $R<5$ kpc. For the GR-1, which lies closer to the Milky Way center, the comparison of the $r$ and $\phi$ direction estimates under the E19 potential suggests that the bulge mass may be smaller than the E19 value of $1.1 \times 10^{10}$ $M_\odot$. For further refining accretion times for these and other phase-space substructures to obtain the accretion history of the Milky Way, more precise astrometric data provided by, e.g., Gaia DR4, China Space Station Survey Telescope and Roman Space Telescope, are needed.

\par All uncertainties reported for the accretion times reflect only statistical errors. The additional systematic effects arising from the adopted potential, both from the choice of the potential and from its intrinsic time evolution, are more substantial and should be addressed in future studies.

\begin{acknowledgments}

\par We thank the anonymous reviewer for the constructive comments and suggestions, which helped improve the clarity and quality of this paper. This work is supported by the National Key Research and Development Program of China No. 2024YFA1611902, National Natural Science Foundation of China (NSFC) No. 12588202, CAS Project for Young Scientists in Basic Research grant No. YSBR-062, the Strategic Priority Research Program of Chinese Academy of Sciences grant No. XDB1160102, and the science research grants from the China Manned Space Project with NO. CMS-CSST-2025-A11. M.C. acknowledges support in part from JSPS KAKENHI (No. JP24K00669 and 25H00394). Supported by International Partnership Program of Chinese Academy of Sciences. Grant No. 178GJHZ2022040GC.

\par This work has made use of data from the European Space Agency (ESA) mission {\it Gaia} (\url{https://www.cosmos.esa.int/gaia}), processed by the {\it Gaia} Data Processing and Analysis Consortium (DPAC, \url{https://www.cosmos.esa.int/web/gaia/dpac/consortium}). Funding for the DPAC has been provided by national institutions, in particular the institutions participating in the {\it Gaia} Multilateral Agreement.

\end{acknowledgments}

\bibliography{Library}
\bibliographystyle{aasjournalv7}


\end{document}